\documentclass[a4paper]{article}
\usepackage[left=20mm, top=20mm, right=20mm, bottom=20mm, nohead]{geometry}
\usepackage[utf8]{inputenc}
\usepackage[american]{babel}
\usepackage{graphicx}
\usepackage{tikz}
\usepackage[colorlinks=true, linkcolor=black, citecolor=black, urlcolor=blue]{hyperref}
\usepackage{cancel}
\usepackage{centernot}
\usepackage{physics}
\usepackage{amssymb}
\usepackage{enumitem}
\usepackage{verbatim}
\usepackage{indentfirst}
\usepackage{simpler-wick}
\usepackage[labelsep=period]{caption}

\allowdisplaybreaks[1]
\usetikzlibrary{decorations.markings}
\tikzset{W->-/.style={decoration={
  markings,
  mark=at position 0.5*\pgfdecoratedpathlength+2pt with
  {\draw[-latex] (-2pt,0pt) -- (1pt,0pt);}},postaction={decorate}},
  W-<-/.style={decoration={
  markings,
  mark=at position 0.5*\pgfdecoratedpathlength with
  {\draw[latex-] (-2pt,0pt) -- (1pt,0pt);}},postaction={decorate}}
  }
\newif\ifWickBelow
\WickBelowfalse
\pgfkeys{
  /simplerwick/.cd,
  arrows/.store in=\LstWickArrows,
  arrows={-,-,-,-,-,-,-,-,-},
  arrows/.initial={-,-,-,-,-,-,-,-,-}, % the # of contractions is bounded by 9
  below/.code={\WickBelowtrue},
}

\makeatletter
\def\swick@end#1#2{
  \swick@setfalse@#1
  \tikzexternaldisable
  \begin{tikzpicture}[remember picture, baseline=(swick-close#1.base)]
    \node[use as bounding box, inner sep=0pt, outer sep=0pt] (swick-close#1) {$\displaystyle #2$};
  \end{tikzpicture}
  \tikz[remember picture, overlay]
{
\foreach \W@X[count=\W@C] in \LstWickArrows
{\ifnum\W@C=#1
\xdef\myW@style{\W@X}
\fi}
\ifWickBelow
    \draw[\myW@style] ($(swick-open#1.south) + (0, -3pt)$)
          -- ($(swick-open#1.base) + (0, -0.5\swick@offset) + #1*(0, -\swick@sep)$)     % Added 0.5 here manually
          -- ($(swick-close#1.base) + (0, -0.5\swick@offset) + #1*(0, -\swick@sep)$)    % Added 0.5 here manually
          -- ($(swick-close#1.south) + (0, -3pt)$);
\else
    \draw[\myW@style] ($(swick-open#1.north) + (0, 3pt)$)
          -- ($(swick-open#1.base) + (0, \swick@offset) + #1*(0, \swick@sep)$)
          -- ($(swick-close#1.base) + (0, \swick@offset) + #1*(0, \swick@sep)$)
          -- ($(swick-close#1.north) + (0, 3pt)$);
\fi}
  \tikzexternalenable}
\makeatother

\graphicspath{{./imgs/}}
\newcommand{\nord}[1]{{:}#1{:}}
\newcommand{\then}{\quad\Rightarrow\quad}

\title{
    \vspace{-2cm} 
    \begin{flushright}{\normalsize INR-TH-2023-005}\end{flushright}
    \vspace{0.5cm}
    Resonant generation of electromagnetic modes \\
    in nonlinear electrodynamics: \\
    Quantum perturbative approach
}

\author{Ilia Kopchinskii$^{1,2}$\thanks{{\bf e-mail}: 	kopchinskii@ms2.inr.ac.ru}$\;$, Petr Satunin$^{2,1}$\thanks{{\bf e-mail}: 	satunin@ms2.inr.ac.ru} \\
\normalsize\it $^1$ Moscow State University, \\ 
\normalsize\it Leninskiye Gory, 119991 Moscow, Russia \\
\normalsize\it $^2$ Institute for Nuclear Research of the Russian Academy
of Sciences, \\  
\normalsize\it 60th October Anniversary Prospect, 7a, 117312  Moscow, Russia}
\date{}

\begin{document}
\thispagestyle{empty}
\maketitle

\begin{abstract}
The paper studies resonant generation of higher-order harmonics in a closed cavity in Euler-Heisenberg electrodynamics from the point of view of pure quantum field theory. We consider quantum states of the electromagnetic field in a rectangular cavity with conducting boundary conditions, and calculate the cross-section for the merging of three quanta of cavity modes into a single one ($3 \to 1$ process) as well as the scattering of two cavity mode quanta ($2 \to 2$ process).
 We show that the amplitude of the merging process vanishes for a cavity with an arbitrary aspect ratio, and provide an explanation based on  plane wave decomposition for cavity modes. Contrary, the scattering amplitude is nonzero for 
 specific cavity aspect ratio.  This $2 \to 2$ scattering is a crucial elementary process for the generation of a quantum of a high-order harmonics  with frequency $2\omega_1 - \omega_2$ in an interaction of two coherent states of cavity modes with frequencies $\omega_1$ and $\omega_2$.
For this process we calculate the mean number of quanta in the final state in a model with dissipation, which supports the previous result of resonant higher-order harmonics generation in an effective field theory approach \cite{OurArticle}.

%, although the process $2 \to 2$ may take place under the coherent initial states in a cavity of 
%This process is related to resonant higher-order harmonics generation in effective field theory approach. 
%Previously, the possibility of resonant amplification has been studied on the classical level in \cite{OurArticle}, where few questions arose: what is the reason for the inhibition of some higher-order harmonics generation; how the generation of $2\omega_1-\omega_2$ harmonic should be understood on quantum level. These questions are successfully answered by the current paper, which applies basics of QFT to the simple 1D- and 3D-cavities in order to calculate mean number of generated signal photons.
\end{abstract}

\section{Introduction}

Nonlinear interaction of electromagnetic field in vacuum, induced by the interaction with virtual electrons, is one of the phenomena theoretically predicted at the dawn of quantum electrodynamics (QED) \cite{Euler:1935zz, Heisenberg:1935qt} but have not been experimentally detected yet due to the weakness of the nonlinear interaction. Specifically, nonlinear dynamics of a strong electromagnetic field is governed by the effective Euler-Heisenberg Lagrangian \cite{Heisenberg:1935qt}. In addition to the QED contribution the theory may obtain correction from hypothetical photophilic scalar and pseudoscalar particles  \cite{Anselm:1985obz,Maiani:1986md,Evans:2018qwy}. 

The variety of phenomena predicted by nonlinear electrodynamics include both perturbative and nonperturbative effects (see e.~g.~\cite{Dunne:2012vv, Dittrich:2000zu, Kuznetsov:2013sea, Fedotov:2022ely} for a review). A significant part of perturbative effects resembles %recalls %reminds
similar effects in nonlinear optical crystals, which include vacuum birefringence of a photon in an external field \cite{Baier:1967zza,Bialynicka-Birula:1970nlh}, and  light-by-light scattering \cite{Euler:1935zz, Heisenberg:1935qt, Rozanov:1993}. The latter effect can be tested at high-intensity laser facilities  (ELI  \cite{ELI}, CoReLS \cite{Corels} etc) by an experiment of three laser beam interaction \cite{Lundstrom-Brodin:2006,Gies-Karbstein:2018,King-Hu-Shen:2018}, see also \cite{Fedotov:2006ii,Sasorov:2021anc, Sundqvist:2023hvw} for related processes. %These processes can be possibly probed at high intensity laser facilities: ELI  \cite{ELI}, CoReLS \cite{Corels} etc.

An alternative approach to test the generation of high-order harmonics is to study this effect for standing electromagnetic modes in high-quality microwave  cavities instead of the traveling waves of lasers \cite{Brodin:2001zz,Eriksson:2004cz}. Although the electromagnetic fields in cavities reach weaker (but still strong) values than in lasers, the signal field may be additionally  enhanced due to the resonant character of the interaction which does not occur in the region of lasers cross-section. Assuming specific set of two pump modes of frequencies $\omega_1$ and $\omega_2$  in the cavity of specific resonant aspect ratio, one obtains resonant enhancement for a mode of frequency $2\omega_1-\omega_2$ \cite{Eriksson:2004cz, Bogorad:2019pbu}. Based on this effect, an experimental proposal aimed on search for nonlinear electrodynamics has been suggested \cite{Bogorad:2019pbu,Kahn:2022uko} which is under design  \cite{Giaccone:2022pke} in Superconducting Quantum Materials and Systems center (SQMS) \cite{SQMS} at Fermi National Accelerator Laboratory.
%the generation of higher-order harmonics. The latter can be tested both with running \cite{Fedotov:2006ii,Sasorov:2021anc, Sundqvist:2023hvw} waves, as well as with standing waves in cavities. The generation of higher-order harmonics for radio modes in cavity may be potentially detected only in case of resonant enhancement %increase
%\cite{Brodin:2001zz}. The 
%solutions for certain sets of pump modes are obtained in \cite{Eriksson:2004cz, Bogorad:2019pbu}, see also \cite{Giaccone:2022pke,Kahn:2022uko} for the experimental proposals. 

Another interesting issue is to study this nonlinear process for an arbitrary set of cavity modes, which can be done analytically for the rectangular shape of the cavity. Thus, when studying nonlinear effects for a single pump mode in one-dimensional cavity it turns out that there is no resonant generation of the third harmonics \cite{Shibata:2020don}. The generalization of the method to the case of two pump modes (frequencies $\omega_1$ and $\omega_2$) and three-dimensional rectangular cavity is given in \cite{OurArticle}. It turns out that the third harmonic as well as the ``plus'' signal mode %(frequency $2\omega_1 +\omega_2$)
are not resonantly amplified in the cavity, the only resonant case is the ``minus'' signal mode of frequency $2\omega_1 -\omega_2$. Aforementioned calculations were performed in the effective field theory approach, in which the classical nonlinear wave equations were solved. It implies that the electromagnetic states are classical states. Remarkably, the final state of higher-order harmonics generation can include %to
very small number of quanta; so one may expect the breakdown of the classical effective field theory approach. Alternatively, one may apply %perform
pure quantum approach based on the calculation of the corresponding matrix elements, which is the purpose of current work. We expect that the quantum calculation will shed more light on the results obtained in the classical calculations.

The difference with the standard quantum perturbative calculation is that the quantum states are eigenstates of the cavity, not the plane waves. The quantization of electromagnetic modes in cavities was previously considered in the context of the Casimir effect \cite{Hacian:1990zw,Hacyan1993}%{\color{blue}+ CITE Shore1991?}
, see also \cite{Chenarani:2013xra} for the symplectic quantization. Although the classical electromagnetic modes relate to the coherent quantum states, we first calculate simple one-particle amplitudes for $3 \to 1$ merging  and $2 \to 2$ scattering processes. We show that $3 \to 1$ merging amplitude vanishes so the third harmonics generation does not occur; $2 \to 2$ scattering, in turn, % performing summation over the coherent states, 
rules the generation of the ``minus'' signal mode.  

The paper is organized as follows: in section \ref{sec:free-quantum} we outline the results of quantization of the free electromagnetic field in rectangular one-dimensional and three-dimensional perfectly conducting cavities. In section \ref{sec:interaction} we include Euler-Heisenberg self-interaction of the electromagnetic field into the Lagrangian and treat it perturbatively with modified $\sf S$-matrix formalism. Several $\sf S$-matrix elements are calculated, which correspond to the key processes discussed on the classical level in \cite{OurArticle}. Finally, the interpretation of results is given in section \ref{sec:conclude}.

\section{Quantum excitations of electromagnetic field in rectangular cavity}
\label{sec:free-quantum}

%Consider free quantum electromagnetic field in a closed cavity $D$,
In this section we consider the quantization procedure for free electromagnetic field in a closed cavity with perfectly conducting boundary conditions, and obtain the spectrum of quantum excitations in one-dimensional and three-dimensional rectangular cavities (cf. \cite{Hacian:1990zw,Hacyan1993,Chenarani:2013xra}). Generally, the free Lagrangian of the electromagnetic field and perfectly conducting boundary conditions read\footnote{Natural system of units $\hbar=c=1$ is assumed.},
%\begin{equation*}
%    \label{eq:free-lagrangian}
%    \mathcal L_0(x) = -\frac{1}{4} \qty\big(F_{\mu\nu}(x))^2, %\qquad t \in\mathbb{R}, ~ \vb r \in D.
%\end{equation*}
\begin{align}
     \label{eq:free-lagrangian}
   & \mathcal L_0 = -\frac{1}{4} \qty\big(F_{\mu\nu})^2,  
\\
\label{eq:boundary-conds}
\eval{\vb E \vdot \vb*{\tau}}_C &= 0, \qquad \eval{\vb B \vdot \vb n}_C = 0, 
\end{align}
%CHANGE #9 in response to Ref.2 Minor comment 1%
where $F_{\mu\nu}$ is a 4-tensor of electromagnetic field, whose components are taken from the strength of electric field $\vb E$ and the force of magnetic field $\vb B$. The perfectly conducting boundary conditions \eqref{eq:boundary-conds} are posed on the projections of field vectors onto the tangent vector $\vb* \tau$ and normal vector $\vb n$ evaluated at the surface $C$ of the cavity.
%END OF CHANGE #9%
%Conducting boundary conditions lead to some differences from the standard case of periodic boundary conditions.

The absence of free charges in the cavity allows us to fix both the Coulomb gauge $\div\vb A(t,\vb r) = 0$, and additionally $A_0 = 0$.
Let us suppose that $\qty{\vb*{\mathcal A}_{n}(\vb r)}$ is a complete orthonormal system of eigenfunctions, satisfying the perfectly conducting boundary conditions.
Here $n$ is an integer number in case of a single finite dimension, or a set of integers in case of larger number of finite dimensions. The vector-potential is decomposed into discrete Fourier series over spatial variable $\vb r$,

%Firstly, we fix Coulomb gauge $\div\vb A(t,\vb r) = 0$, where $\vb A(t,\vb r)$ --- electromagnetic vector-potential. The absence of free charges in the cavity allows us additionally to fix the gauge, $A_0 = 0$. Let the cavity have a complete orthonormal system of eigenfunctions $\qty{\vb*{\mathcal A}_n(\vb r)}, ~ n \in \mathbb{N}$, satisfying the electric boundary conditions of perfectly conducting walls. The vector-potential is then decomposed into discrete Fourier series over spatial variable $\vb r$ (translational symmetry broken) and Fourier integral over time (temporal symmetry preserved). The integral is conventionally cancelled to positive and negative frequency components with the help of Dirac delta expressing the dispersion relation,
\begin{equation}
    \label{eq:free-expand}
    \vb A(t,\vb r) = \vb A^+(t,\vb r) + \vb A^-(t,\vb r), 
    \qquad \vb A^\pm(t,\vb r) = \frac{1}{\sqrt{V}} \sum\limits_{n} a^\pm_n \vb*{\mathcal A}_n(\vb r) \frac{e^{\pm i\omega_n t}}{\sqrt{2\omega_n}},
\end{equation}
where $V$ is the cavity's volume. The dimensionless amplitudes $a^\pm_n$ in the quantization procedure become the creation and annihilation operators $\hat{a}^\pm_n$ satisfying commutation relations $\comm{\hat{a}^-_n}{\hat{a}^+_k} \equiv \delta_{nk}$. 
The operators $\hat{a}^\pm_n$ act on the Fock space containing the vacuum state $\ket{0}$, which is vanished by acting the annihilation operator $\hat{a}^-_n \ket{0}=0$ for arbitrary $n$; $m$-particle states $\ket{m_n}$ are built with $a^+_n$ acting on the vacuum $\ket{0}$, $\ket{m_n} = \frac{\qty(a_n^+)^m}{\sqrt{m!}} \ket{0}$. Classical electromagnetic waves are related to the coherent states $\ket{\xi_n}$ defined as $\ket{\xi_n} = e^{-\frac{\abs{\xi}^2}{2}} e^{\xi a^+_n} \ket{0}$.
%\begin{align}
%    \label{eq:pure}
%    \qq*{Pure state with exactly $\nu$ photons:} \ket{\nu_n} &= \frac{\qty(a_n^+)^\nu}{\sqrt{\nu!}} \ket{0}, \qquad \nu \in \mathbb{N}_0,
%    \\
%    \label{eq:coherent}
%    \qq*{Coherent state with $\abs{\xi}^2$ photons on average:} \ket{\xi_n} &= e^{-\frac{\abs{\xi}^2}{2}} e^{\xi a^+_n} \ket{0}, \qquad \xi \in \mathbb{C}.
%\end{align}

\subsection{One-dimensional cavity}
\label{sec:free-1d}
Let us consider free electromagnetic field confined in an effective one-dimensional cavity $D = (0,L_x)$ between two large perfectly conducting planes of area $S = L_y L_z$ such that $\sqrt{S} \gg L_x$, and volume $V = L_x S$.
%\begin{figure}[ht!]
%\vspace{-8mm}
%\centering
%\includegraphics[width=\linewidth]{Linear-cavity-A.png}
%\caption{Vector-potential for $n=1$ cavity mode for 1D cavity with perfectly conducting walls. Two components of vector-potential relate to two linear polarizations.}
%\label{img:linear-A}
%\end{figure}
The vector-potential depends only on $(t,x)$ coordinates, so that condition $\div{\vb A} = 0$ implies $\vb A \perp \vb e_x$. % so that $\vb E$.
Electric boundary conditions $\eval{\vb E \vdot \vb e_x}_{x=0,L_x} = 0$ translate literally for the vector-potential, which is therefore decomposed into sine eigenfunctions,
\begin{equation}
    \label{eq:free-1d}
    A_i^\pm(t,x) = \sqrt{\frac{2}{V}} \sum\limits_{n=1}^\infty a^\pm_{i,n}\sin(k_n x) \frac{e^{\pm i \omega_n t}}{\sqrt{2\omega_n}},
    \qquad
    k_n = \omega_n = \frac{\pi n}{L_x},
    \qquad
    i \in \{y,z\}.
\end{equation}
Two independent ($y$ and $z$ directed) components are related to two linear polarizations.
%Note that $y-$ and $z-$ components are independent, and related to two linear polarizations.
%Any polarization of the electromagnetic field can be expressed in terms of linear polarizations along two orthogonal directions $\vb e_y$ and $\vb e_z$. % Eventually, the amplitudes $a^\pm_{i,n}$ are quantized to ladder operators as usual.
Any one-particle state for fixed $n$ in one-dimensional cavity is expressed as a linear combination of two orthogonal states with $y$ and $z$ linear polarizations, $\ket{1_{n,y}} = a^+_{n,y}\ket{0}$ and  $\ket{1_{n,z}} = a^+_{n,z}\ket{0}$. %Generally, a state from the Fock space decomposes into states with $N_y$ particles of y-polarization and $N_z$ -- of z-polarization, $\ket{N^{y}_}$ 

Note that the cavity modes are not the eigenstates of the spatial momentum operator. However, they can be easily decomposed to two plane waves traveling in opposite directions which are the spatial momentum eigenstates, $\ket{1_n}= \frac{1}{2i}\qty(\ket{k_n} - \ket{-k_n})$ (polarization index is omitted). Here $\ket{k_n}$ denote the state with definite momentum $(k_n,0,0)$. In turn, the plane waves $\ket{k_n}$ do not satisfy the conducting boundary conditions so cannot be considered as the cavity eigenstates.

% \begin{gather*}
%     \div\vb E = \partial_x \partial_t A_x(t,x) + \partial_y \partial_t A_y(t,x) + \partial_z \partial_t  A_z(t,x) = 0 \then \partial_t \partial_x A_x(t,x) = 0,
%     \\
%     \partial_t \partial_x A_x(t,x) = 0 \then A_x(t,x) = c_0 + c_1 t + c_2 x \then E_x(t,x) = c_1.
% \end{gather*}
% \begin{equation*}
%     x = 0, L_x: \qquad \vb E \parallel \vb n \quad\Leftrightarrow\quad \partial_t \vb A_\perp = 0 \quad\Leftarrow\quad A_y = 0, ~ A_z = 0.
% \end{equation*}

\subsection{Rectangular cavity}
\label{sec:free-3d}

The following case is a three-dimensional rectangular cavity of dimensions $L_x, L_y, L_z$  with perfectly conducting walls. The volume of the cavity is $V = L_x L_y L_z$.
%\begin{figure}[ht!]
%\vspace{-8mm}
%\centering
%\includegraphics[width=\linewidth]{Rectangular-cavity.png}
%\caption{3D cavity with perfectly conducting walls.}
%\label{img:box-cavity}
%\end{figure}
The electromagnetic potential %$\vb A_\mu$ 
$\vb A$ decomposes into cavity eigenmodes $\vb*{\mathcal A}^\lambda_{npq}(\vb r)$  which are conventionally grouped into TE and TM modes,
%The eigenfunctions are TM- and TE-modes $\vb*{\mathcal A}^\lambda_{npq}(\vb r)$ which the spatial part of the electromagnetic vector-potential is decomposed into:
\begin{equation}
    \label{eq:free-3d}
    \vb A^\pm(t,\vb r) = \sqrt{\frac{1}{V}} \sum\limits_{\lambda,npq}^\infty a^{\lambda\pm}_{npq} \vb*{\mathcal A}^\lambda_{npq}(\vb r)  \frac{e^{\pm i \omega_{npq} t}}{\sqrt{2\omega_{npq}}},
    \qquad
    \omega_{npq} = \sqrt{\frac{\pi^2 n^2}{L_x^2} + \frac{\pi^2 p^2}{L_y^2} + \frac{\pi^2 q^2}{L_z^2}},
    \qquad
    \lambda \in \{\text{TM}, \, \text{TE}\}.
\end{equation}
Here $n,p,q$ are integer numbers related to the harmonics number over each spatial dimension. The spectrum $\omega_{npq}$ is degenerate if all $n,p,q$ are nonzero: $\mbox{TE}_{npq}$ and $\mbox{TM}_{npq}$ are two independent solutions related to the same eigenfrequency, so the TE/TM property may be treated as a polarization basis. 
The orthonormal eigenfunctions of the rectangular cavity $\vb*{\mathcal A}^\lambda_{npq}(\vb r)$ are proportional to the dimensionless eigenfunctions of the electric field, see \cite{Hill:2014},
\begin{equation*}
    \label{eq:eigen-3d}
    \scriptsize
    \vb*{\mathcal{A}}^\text{TM}_{npq}(\vb r) = \frac{\sqrt{4(2-\delta_{q0})}}{\omega_{npq}}
    \begin{pmatrix}
        \dfrac{k_x k_z}{\sqrt{k_x^2+k_y^2}} \cos(k_x x)\sin(k_y y)\sin(k_z z) \\
        \dfrac{k_y k_z}{\sqrt{k_x^2+k_y^2}} \sin(k_x x)\cos(k_y y)\sin(k_z z) \\
        -\sqrt{k_x^2+k_y^2}\sin(k_x x)\sin(k_y y)\cos(k_z z)
    \end{pmatrix}\!\!,
  \  % \qquad
    \vb*{\mathcal{A}}^\text{TE}_{npq}(\vb r) = \! \sqrt{4(2-\delta_{n0}-\delta_{p0})}
    \begin{pmatrix}
        +\dfrac{k_y}{\sqrt{k_x^2+k_y^2}} \cos(k_x x)\sin(k_y y)\sin(k_z z) \\
        -\dfrac{k_x}{\sqrt{k_x^2+k_y^2}} \sin(k_x x)\cos(k_y y)\sin(k_z z) \\
        0
    \end{pmatrix},
\end{equation*}
where $k_x(n) = \frac{\pi n}{L_x}$, $k_y(p) = \frac{\pi p}{L_y}$ and $k_z(q) = \frac{\pi q}{L_z}$. 
%The modes are numbered with three indexes, which run slightly different sets of values for TM- and TE-modes.
Here the normalization factors $\sqrt{4(2-\delta_{q0})}$ and $\sqrt{4(2-\delta_{n0}-\delta_{p0})}$ are introduced to satisfy the normalization condition $\norm{\vb*{\mathcal{A}}^\lambda_{npq}}^2 = V$ for zero and nonzero wavenumbers simultaneously. %keep the uniformity of the eigenfunctions whenever zero indexes occur.
%Two discrete degrees of freedom, being intrinsic to the electromagnetic field, express in two possible values for the polarization index $\lambda \in \{\text{TM}, \, \text{TE}\}$.
One-particle states are produced by the action of the creation operator, $\ket{1^\lambda_{npq}}=a^{\lambda +}_{npq}\ket{0}$. 
Similar to the one-dimensional cavity case, the rectangular cavity eigenstates decompose to finite (now $8$) number of plane waves.

\section{Transition amplitudes in the presence of the electromagnetic field self-interaction}
\label{sec:interaction}

In this section we discuss the electromagnetic field self-interaction, and calculate certain transition amplitudes between cavity quantum states. The self-interaction of the electromagnetic field is described by a quartic nonlinear Lagrangian, %   that can be decomposed as follows $\mathcal L_\text{EH} = \mathcal L_0 + \Delta \mathcal L_\text{EH}$, where  
\begin{equation}
    \label{eq:EH-term}
   % \mathcal L_0 = -\frac{1}{4} \mathcal F,
   % \qquad
    \mathcal L_\text{int} = \kappa\left((F_{\mu\nu}F^{\mu\nu})^2 + \beta(F_{\mu\nu}\tilde{F}^{\mu\nu})^2\right) = 4\kappa\qty(\vb E^4 - 2\vb B^2 \vb E^2 + \vb B^4 + 4\beta(\vb B\vb E)^2).
\end{equation}
The values $\kappa = \frac{\alpha_e^2}{90 m_e^4}$ and $\beta=\frac{7}{4}$ related to the standard Euler-Heisenberg Lagrangian \cite{Heisenberg:1935qt} are obtained by integrating out electrons in pure QED. Nevertheless, we keep arbitrary values of $\kappa$ and $\beta$ in order to take into account possible new physics contribution.  
The calculation of certain transition amplitudes $\braket{f(t_f)}{i(t_i)}$ between cavity eigenstates $\ket{i}$ and $\ket{f}$ for a time interval $(t_f-t_i)$ can be performed in the perturbative approach assuming $ \mathcal L_\text{int}$ to be small. The transition probability $w_{fi}$ is a squared modulus of the amplitude, $w_{fi} (t_f-t_i) = |\braket{f(t_f)}{i(t_i)}|^2$.

%and apply perturbative (over small $\kappa \mathcal F$) approach to calculation of certain $\sf S$-matrix elements $\mel{f}{\sf S}{i}$ between cavity eigenstates.

An essential difference from the standard perturbation theory in relativistic quantum field theory (QFT) is that the initial and final states are no longer plane waves but cavity modes captured in a closed volume. Thus, one cannot straightly define asymptotic states at spatial infinity which are necessary for the standard S-matrix formalism. %, and construct the standard $\sf S$-matrix formalism.
In order to solve it one can either construct the perturbation theory for cavity amplitudes from the first quantum mechanical principles or try to reduce the task to the standard one. The latter approach is simpler, so we apply it assuming $t_{f(i)} \to +(-) \infty$.
%CHANGE #4 in response to Ref.2 Major comment 2%
As long as the nonlinear Euler-Heisenberg term \eqref{eq:EH-term} remains small in comparison to the classical linear term \eqref{eq:free-lagrangian}, i.e. $\kappa \vb B^4 \ll \vb B^2$ or $\abs{\vb B} \ll m_e^2 / \alpha_e \sim 10^9$ T,

% Add smth about 'highly confined geometries'?
%END OF CHANGE #4%
we can decompose the cavity modes as linear combinations of plane waves $\ket{i} = \Sigma_n c^i_n\ket{k_n}$\footnote{Here $n$ is a single integer in case of a single closed dimension, and a set of integers in case of more closed dimensions.}. Then the transition amplitude $\braket{f(t_f)}{i(t_i)}$ is expanded into the linear combination of transition amplitudes between plane wave states which are expressed as $\sf S$-matrix elements, % between plane wave states,
\begin{equation}
    \braket{f(t_f)}{i(t_i)} = \Sigma_{nm} (c^f_m)^*c^i_n \mel{k_m}{\sf S}{k_n}.
    \label{Smx}
\end{equation}
The following calculations may %be
proceed in two ways. First, one can calculate plane wave amplitudes and take a sum (\ref{Smx}) over them with the appropriate coefficients. The complexity of the summation quadratically grows with the number of plane waves in the decomposition. Alternatively, due to the linearity of (\ref{Smx}) with respect to the states, the sums can be taken forming the initial and final states back, $\mel{f}{\sf S}{i}$; the $\sf S$-matrix and the corresponding matrix element read the standard form,
%Due to the extreme smallness of $\kappa \mathcal F \lesssim 10^{-24}$ (under laboratory conditions), we employ perturbative approach. 
%Since eigenfunctions of rectangular cavities decompose into the plane waves, the generic formalism of $\sf S$-matrix is applicable to describe transitions between field states discussed above. Therefore, we calculate $\sf S$-matrix elements $\mel{f}{\sf S}{i}$ whose squared modulo is proportional to the probability of the transition $\ket{i} \rightarrow \ket{f}$,
\begin{equation}
    \label{eq:S-matrix}
    {\sf S} = \mathrm{Texp}\,\qty(i\int\nord{ \mathcal L_\text{int}(x)} \dd[4]{x})
    \then
    \mel{f}{\sf S}{i} = i\int \mel{f}{\nord{ \mathcal L_\text{int}(x)}}{i} \dd[4]{x} + \order{\kappa^2},
\end{equation}
where we expand the T-exponent up to the first non-vanishing term $\propto \kappa$. In order to perform the calculation, one needs to determine the Wick's contractions of ladder operators and electromagnetic fields, for one-dimensional cavity,
\begin{equation}
    \label{eq:wick-1d}
    \wick[below]{\c a^-_{i,n} \, \c E_j(t,x)} = \delta_{ij} \, i \sqrt{\frac{\omega_n}{V}} \sin(k_n x) \, e^{i\omega_n t},
    \qquad
    \wick[below]{\c a^-_{z,n} \, \c B_y(t,x)} = -\sqrt{\frac{\omega_n}{V}} \cos(k_n x) \, e^{i\omega_n t} = -\wick[below]{\c a^-_{y,n} \, \c B_z(t,x)},
\end{equation}
and for rectangular cavity,
\begin{equation}
    \label{eq:wick-3d}
    \wick[below]{\c a^{\lambda-}_{npq} \, \c {\vb E}(t,\vb r)} = i \sqrt{\frac{\omega_{npq}}{2V}} \vb*{\mathcal A}^\lambda_{npq}(\vb r) \, e^{i\omega_{npq} t},
    \qquad
    \wick[below]{\c a^{\lambda-}_{npq} \, \c {\vb B}(t,\vb r)} = \frac{1}{\sqrt{2\omega_{npq}V}} \curl\vb*{\mathcal A}^\lambda_{npq}(\vb r) \, e^{i\omega_{npq} t}.
\end{equation}
These contractions can be straightforwardly verified by substituting electric and magnetic fields' expansions (based on \eqref{eq:free-1d}, \eqref{eq:free-3d}) into the l.h.s of \eqref{eq:wick-1d}, \eqref{eq:wick-3d}.% and applying there the definition of ladder operators $\wick[below]{\c a^-_n \, \c a^+_m} = \comm{a^-_n}{a^+_m} \equiv \delta_{nm}$.

Computing the $\sf S$-matrix element (\ref{eq:S-matrix}) with the Wick contractions given in (\ref{eq:wick-1d}), (\ref{eq:wick-3d}), one integrates over the four-dimensional volume. The spatial integration $\int d^3x$ is carried out over the cavity modes in the volume of the cavity; the time integration is carried out over the time oscillating exponents for initial and final states, $\int_{-\infty}^{+\infty} dt\; e^{i(\Sigma\omega_i-\Sigma\omega_f)t} = 2\pi \delta(\Sigma\omega_i-\Sigma\omega_f)$,  giving the delta-function which represents the energy conservation. In the standard scattering problem (see a textbook \cite{Schwartz:2014sze}) the factor $2\pi \delta(0)$ is interpreted as the time of the interaction $T$ between wavepackets which goes to infinity in case of plane waves; the scattering probability grows linearly with $T$.

Two changes appear in our task. First, the interaction time is finite and should be bounded by a coherence time,  $\int_{t_{coh}} dt\; e^{i(\Sigma\omega_i-\Sigma\omega_f)t} = t_{coh}$ if $\Sigma\omega_i=\Sigma\omega_f$. Second, one should also take into account dissipation which is introduced %taken into account by the addition of
as a small imaginary additive to frequency, $\omega_f - i \frac{\omega_f}{Q}$. Thus, the time integral gives $\max\qty(t_{coh},\frac{Q}{\omega_f})$.
%CHANGE #5 in response to Ref.2 Major comment 3%

This treatment of dissipation is valid in the limit of high quality factor of the cavity $Q \gg 1$. It ceases to be applicable when the field cannot be quantized as a set of harmonic oscillators \eqref{eq:free-expand}, since interaction with the walls (being source of dissipation) makes the oscillations sufficiently an-harmonic. In general case the dissipation can be treated rigorously on quantum level, e.g. within the non-equilibrium diagram technique, %or maybe \cite{Deyo-Kelley:2022}?%
although we restrict our treatment to the case $Q \gg 1$.
%END OF CHANGE #5%

%Since the cavity amplitude decomposition to the linear combination of the scattering ones  holds, aforementioned interpretation transfers as a whole to our task:  $2\pi \delta(0)$ is to be replaced with the interaction time $t$.
%Introducing more realistic dissipation case to our task,  the replacement $2\pi \delta(0) \to Q/\omega $, where $Q$ is the quality factor, should be done.

%In the process of the evaluation of the matrix element $\mel{f}{\sf S}{i}$  the infinite factor $2\pi \delta(0)$ appears. Introducing dissipation, it becomes to $2\pi /Q$.

\subsection{The merging process \texorpdfstring{$3 \to 1$}{3 --> 1} in one-dimensional cavity}
\label{sec:proc-3-1}
%In the classical approach \cite{OurArticle} the monochromatic excitation of the linear cavity was carried out foremost. A counter-intuitive result read that generation of the third harmonic is impossible in such layout. Let us explain this fact on the quantum level.

Let us first consider the linear cavity, and calculate the amplitude for the merging of three quanta of pump modes into a single quantum of signal mode. The cases of three quanta of the same pump mode, as well as two quanta of one pump mode $\omega_1$ and a quantum of a mode $\omega_2$ can be considered simultaneously. Additionally, any quantum state may be taken of different polarization. 
%energy $\omega_n$ into a single quanta of energy $\omega_{3n}=3\omega_n$ (the energy conservation holds, see eq.(\ref{eq:free-1d})).   
%The merging of three photons $\omega_n$ into a single photon $\omega_{3n}$ (denoted $3\omega_n \longrightarrow \omega_{3n}$) is not prohibited by the law of energy conservation. Therefore it is adequate to choose 
Generally, the initial and final quantum field states read,
\begin{equation}
    \label{eq:states-21-1}
    \ket{i} = \ket{1^i_n} \otimes \ket{1^j_n} \otimes \ket{1^l_p} = a^+_{i,n} a^+_{j,n} a^+_{l,p} \ket{0},
    \qquad\qquad
    \ket{f} = \ket{1^s_{2n+p}} = a^+_{s,2n+p} \ket{0}.
\end{equation}
The indices $n$ and $p$ relate to wavenumbers of two pump modes ($n=p$ means the case of single pump mode); the indices $i,j,l,s$ relate to polarization for every modes.
%constructed with the ladder operators mentioned in subsection \ref{sec:free-1d}. We keep arbitrary polarization indexes $i,j,l \in \{y,z\}$ for completeness.
The calculation of the $\sf S$-matrix element proceeds as follows:
\begin{equation}
  \mel{f}{\sf S }{i} = i \kappa \int\limits_{-\infty}^{+\infty}\dd{t} \iint\dd{S} \int\limits_0^{L_x}\dd{x} \mel{f}{\nord{(\vb E \vb E)^2 - 2\vb B^2 \vb E^2 + (\vb B \vb B)^2 + 4\beta(\vb B\vb E)^2} }{i},
   \label{Tfi}
\end{equation}
where $ \iint\dd{S}=S$ denotes the area of transverse cavity dimensions. %the squares of the field invariants $\mathcal F^2, \mathcal G^2$ can be  expressed in terms of electric and magnetic field components (convenient for the evaluation of Wick contractions and for the integration over cavity length). The integral over time traditionally collapses into the factor $2\pi\delta(0)$ which will be interpreted later. %Note that the terms with $\expval{\mathcal F^2}$ and $\expval{\mathcal G^2}$ are treated independently. 
%Let us rewrite the invariant in terms of E and B,
%\begin{equation*}
%    \label{eq:averaging-1d}
%    \mel{f}{\sf S}{i} = \expval{\mathcal F^2} + \beta \expval{\mathcal G^2}, \qq{where} \expval{F(x)} = i\kappa S 2\pi\delta(0) \int\limits_0^{L_x}\dd{x} \mel{0}{a^-_{l,3n} \nord{F(x)} \qty(a^+_{i,n})^2 a^+_{j,n}}{0}.
%\end{equation*}
%\begin{equation}
%\label{Tfi}
% {\sf T}_\text{fi} &= 4i\kappa S \int\limits_{-\infty}^{+\infty}\dd t \int\limits_0^{L_x}\dd x\expval{a^-_{s,2n+p} \nord{\vb E^4 - 2\vb B^2 \vb E^2 + \vb B^4 + 4\beta(\vb B\vb E)^2} a^+_{i,n} a^+_{j,n} a^+_{l,p}}{0} .
%\end{equation}
Each term in (\ref{Tfi}) reads (see Appendix \ref{app:details-3-1} for details), 
\begin{equation}
\expval{(\vb E \vb E)^2} = \expval{(\vb B \vb B)^2} = -\frac{1}{2}\expval{ \vb B^2 \vb E^2} = %2\pi\delta(0) 
\frac{\sqrt{(2n+p)n^2p}\pi^2}{L_x^3} \qty[\delta_{ij}\delta_{ls} (1 + 2\delta_{is}) + (1-\delta_{ls}) (1-\delta_{ij})],
\end{equation}
and $ \langle(\vb E \vb B)^2\rangle = 0$. Here no summation over polarization indices $i,j,l,s$ is assumed.
The brackets  $\langle ...\rangle$ denote the integral over $dx$ for the matrix element of the corresponding operator. Taking all the contributions together we obtain that the amplitude vanishes, $ \mel{f}{\sf S }{i}=0$, for any $\beta$: the amplitude components related to two electromagnetic invariants vanish independently, 
$  \expval{(F_{\mu\nu})^4} = \expval{( F_{\mu\nu}\tilde{F}^{\mu\nu})^2} = 0$.
% It turns out that the matrix element component related to of each squared field invariant vanishes independently, % The $\sf S$-matrix element:

%We find that the matrix element component related to of each squared field invariant vanishes independently, without the need to compensate each other. 
Hence, neither a monochromatic excitation of linear cavity $\omega$ cannot generate the third harmonic $3\omega$ nor two pump modes $\omega_1$ and $\omega_2$ cannot merge into a mode $2\omega_1 + \omega_2$ for any polarization of incoming photons. %The process $3 \rightarrow 1$ is  prohibited not by the energy conservation but by the internal structure of interaction Lagrangian. 
More insight for the reason of this prohibition can be obtained with the plane wave decomposition.

\paragraph{Plane wave decomposition.} Let us provide a more clear interpretation for the reason of non-generation of the third harmonics or ``plus'' combined signal mode. The amplitude $\mel{f}{\sf S}{i} = \mel{1^s_{2n+p}}{\sf S}{1^i_n,1^j_n,1^l_p} $ decomposes (see Sec.~2.1) into 16 plane wave amplitudes, 
\begin{equation}
\label{Sfi_plane_wave}
\mel{f}{\sf S}{i}  = \sum_{(\pm)^4} \frac{(-1)^\pm}{(2i)^4} \mel{\pm k^s_{2n+p}}{\sf S}{\pm k_n^i, \pm k_n^j,\pm k_p^l},    
\end{equation}
where $(-1)^\pm = 1$ in case of even number of signs ``$+$'' in the notations for all initial and final plane wave states, and $-1$ otherwise. The momentum conservation requires (see eq.~\ref{eq:free-1d}) that all signs in the states notation should coincide, ``$+$'' or ``$-$'' simultaneously. Hence, the only remaining nonzero terms in (\ref{Sfi_plane_wave}) are 
\begin{equation}
\mel{f}{\sf S}{i}  = \frac{1}{(2i)^4} \mel{k^s_{2n+p}}{\sf S}{ k_n^i, k_n^j, k_p^l} + \frac{1}{(2i)^4} \mel{-k^s_{2n+p}}{\sf S}{ -k_n^i, -k_n^j, -k_p^l} =  \frac{2}{(2i)^4} \mel{k^s_{2n+p}}{\sf S}{ k_n^i, k_n^j, k_p^l}.
\end{equation}
%shows that 14 amplitudes are zero, the remaining ones are $ \frac{1}{(2i)^4} \mel{k^s_{2n+p}}{\sf S}{ k_n^i, k_n^j, k_p^l}$ and  $\frac{1}{(2i)^4} \mel{-k^s_{2n+p}}{\sf S}{ -k_n^i, -k_n^j, -k_p^l}$. 
%These amplitudes are equal and, in turn, vanish since the Lorentz-invariant matrix element should contain only the scalar products of momenta and polarization vectors which vanish at the kinematic configuration of all parallel momenta.
Thus, the 4-point amplitude describing $3 \to 1$ merging process for cavity eigenstates reduces to the corresponding amplitude for plane waves; all plane wave momenta must be parallel. The plane wave matrix element, in turn, satisfy Lorentz Invariance. Hence, the amplitude should contain only Lorentz scalars which can be only the scalar products of momenta and polarization vectors. For massless transverse photons all these scalar products vanish at the kinematic configuration of parallel momenta.

\paragraph{Plane wave decomposition for three-dimensional cavity.} Aforementioned consideration can be in a direct way generalized to the case of $3 \to 1$ merging process  $\mel{1^s_{(2n+n'),(2p+p'),(2q+q')}}{\sf S}{2^\lambda_{npq},1^{\lambda'}_{n'p'q'}} $\footnote{Both the third harmonics generation ($n'=n, p'=p, q'=q$) and the generation for a mode of combined frequency $2\omega_1 + \omega_2$ are considered.} in %the third harmonics generation by a single pump mode in 
three-dimensional rectangular cavity. %both for the third harmonics generation, and the generation for a mode of combined frequency $2\omega_1 + \omega_2$. 
In fact, although the plane wave decomposition give $8^4$ plane wave amplitudes, due to the momentum conservation the only nonzero terms contain only the states of parallel momenta, which in turn vanishes due to the reasons given in the previous paragraph.  %as well for the case of two pump modes with parallel momenta. % We hope that the same reason works for the 3rd harmonics generation in 3D, as well as for the case of parallel $(2n+p)$ in 3D.

\subsection{The scattering process $2\to 2$ in three-dimensional cavity} %``minus'' signal mode generation}
\label{sec:coh-gen}

The classical  EFT analysis \cite{OurArticle}, supported with the quantum calculation of Sec.~3.1, shows that in three-dimensional cavity with pump modes of frequencies $\omega_1$ and $\omega_2$ there is no resonant generation of signal modes $3\omega_1$ ($3\omega_2$) and $2\omega_1 + \omega_2$ ($2\omega_2 + \omega_1$). However, in case of special resonant cavity geometry there is a classical solution describing resonant generation of the mode $2\omega_1-\omega_2$. 
The energy conservation shows that the energy of a final quantum in the merging process of three quanta into a single one is a sum of incoming quanta energies $2\omega_1 + \omega_2$. 
%Although generation of higher-order harmonics $3\omega_1$ and $2\omega_1 + \omega_2$ does not take place in rectangular cavities, the classical analysis \cite{OurArticle} expects resonant amplification of $2\omega_2-\omega_1$ in the 3D box cavity. Namely, the generation of TM130 signal mode from pump modes TM110 and TE011 is approved for certain cavity geometry $L_x = L_y = L_z/r$, where $r = \qty[\sqrt{5}-2]^\frac{1}{2}$, so that $\omega_{130} = 2\omega_{011} - \omega_{130}$.
%The naive quantum description of such process fails, the only possible result of such process is the energy of final quanta $2\omega_1 + \omega_2$ due to the energy conservation, while the classical approach lead to the energy $2\omega_1 - \omega_2$
This apparent contradiction can be explained as follows.

%However, the naive quantum description of such process fails, since the synthesis $\omega_{110} + 2\omega_{011} \centernot{\longrightarrow} \omega_{130}$ is prohibited by the energy conservation law. Instead the scattering $2\omega_{011} \longrightarrow \omega_{130} + \omega_{110}$ is allowed, although the classical calculation shows that monochromatic generation (with a single TE011 pump mode) is still impossible.

% Разногласие классики и квантов мнимое, но хочется как-то назвать описанное затруднение, чтобы решаемая нами трудность ощущалась весомой. 

The precise quantum correspondent of a classical wave is a coherent state which is an eigenstate of the electromagnetic field operator but not an eigenstate of operator of a particle number. Hence, the full number of quanta in a process involving the coherent states in initial and final states is not fixed. % A coherent state decomposes to fixed number states with coefficients equal to square root of Poisson distribution. Number of particles is not conserved. 
Since, one may suppose that the $2\omega_1-\omega_2$ signal mode generation at a quantum one-particle level is a result of $2 \to 2$ ``scattering process'' instead of $3 \to 1$ merging. The only option for this elementary $2 \to 2$ process, satisfying the energy conservation, is a scattering of two quanta of the pump mode $\omega_1$ to a single quantum of mode $\omega_2$ and a quantum of the signal mode $2\omega_1-\omega_2$. The energy conservation holds $2\omega_1 \to \omega_2 + (2\omega_1 - \omega_2)$.  % The number of quanta of $\omega_2$ mode in the final state increases, but there is no contradiction.

Let us perform the calculation of $2 \to 2$ for a particular configuration with a simple set of modes previously considered in classical effective theory approach in \cite{OurArticle}, precisely scattering of $2$ quanta TE011 to a quantum of TM110 and a quantum of TM130. The energy conservation reads,
\begin{equation}
\label{energycons}
2\omega_{011} = \omega_{110} + \omega_{130}.
\end{equation}
Considering cavity of dimensions $L_x:L_y:L_z=1:1:r$ and taking into account the expressions (\ref{eq:free-3d}) for given mode frequencies
\begin{equation}
\label{omegas}
\omega_{011} = \frac{\pi}{L_z}\sqrt{r^2 + 1}, \qquad \omega_{110} = \frac{\pi}{L_z}\sqrt{2}\,r, \qquad \omega_{130} = \frac{\pi}{L_z}\sqrt{10}\,r,    
\end{equation}
one obtains $r = \sqrt{\sqrt{5}-2}$, which %coincides with 
is identical to the resonant condition in the effective theory approach \cite{OurArticle}. 
%CHANGE #3 in response to Ref.2 Major comment 1%
 It is worth mentioning that the resonant condition can be satisfied in other rectangular configurations where $L_x \neq L_y$, the choice \eqref{omegas} is made only for simplicity. The general analysis of resonant conditions for a rectangular cavity, performed in \cite{OurArticle}, suggests that resonance is not prohibited for many other combinations of pump and signal modes, the selected combination of TE011, TM110 and TM130 being just a particular example.
% About other geometries: has anybody found resonant combinations in cylidrical cavity? Or in spherical cavity?
%END OF CHANGE #3%

%Let us check it with a direct quantum calculation for the TM130 signal mode production by the TE011 + TM110 combination of pump modes, described in classical effective theory in \cite{OurArticle}.
%The resonance for the mode frequencies $\omega_{TM130} = 2\omega_{TE011}-\omega_{TM011}$ lead to the following resonant ratio of the cavity dimensions, \\ $L_x:L_y:L_z=1:1:r$, where $r = \left({\sqrt{5}-2}\right)^{1/2}$. Exactly the same ratio obtained in quantum approach from the energy conservation, $ \Sigma\omega_i -\Sigma \omega_f = 2\omega_{TE011} - \omega_{TM130} - \omega_{TM011} = 0$. %By the way, we will not use the numerical value of $r$ during the calculations, and obtain the resonant condition for $r$ independently.

The initial and final states for the $2 \to 2$ process read,
\begin{equation}
     \ket{i} = \ket{2^\text{TE}_{011}} = \frac{1}{\sqrt{2}}\left(a^{\text{TE}+}_{011}\right)^2\ket{0},  \qquad   \ket{f} = \ket{1^\text{TM}_{110}} \otimes \ket{1^\text{TM}_{130}} = a^{\text{TM}+}_{110}a^{\text{TM}+}_{130}\ket{0}.
\end{equation}
The matrix element of the $2 \to 2$ process $\mel{f}{\sf S}{i} \equiv 2\pi i \delta \left(\Sigma\omega_i - \Sigma\omega_f\right) {\sf M}_{2\to 2}$ reads,
\begin{equation}
\label{eq:mel-2-2}
{\sf M}_{2\to 2} = 4\kappa \int\limits_V\dd^3x \mel{f}{\,\nord{\vb E^4 - 2\vb B^2 \vb E^2 + \vb B^4 + 4\beta(\vb B\vb E)^2}\,}{i} = \expval{\vb E^4} - 2\expval{\vb B^2\vb E^2} + \expval{\vb B^4} + 4\beta\expval{(\vb B\vb E)^2},
\end{equation}
where the brackets around an operator imply the
corresponding part of the matrix element ${\sf M}_{2\to 2}$.
%taken matrix element  and integration over the cavity volume.
%\begin{equation}
% \expval{F} = \int\limits_V\dd^3 x \expval{a^{\text{TM}-}_{130} a^{\text{TM}-}_{110} \, \nord{F} \, a^{\text{TE}+}_{011} a^{\text{TE}+}_{011} }{0}.
% \end{equation}
Performing the Wick contractions (\ref{eq:wick-3d}) and integrating over the cavity volume one obtains (see Appendix \ref{app:details-2-2} for details),
\begin{align}
&\expval{\vb E^4} =  -\frac{4\kappa}{\sqrt{2}V}\sqrt{\omega_{011}^2\omega_{110}\omega_{130}}, \qquad \qquad
\expval{\vb B^4} = \frac{4\kappa}{\sqrt{2}V} \frac{\pi^4}{L_z^4}2r^2\frac{ 2r^2 - 3}{ \sqrt{\omega_{011}^2\omega_{110}\omega_{130}}}, \label{22-1}\\
&\expval{\vb B^2 \vb E^2} =\frac{4\kappa}{2\sqrt{2}V}\frac{\pi^2}{L_z^2}\left(4r^2 \sqrt{\frac{\omega_{011}^2}{\omega_{110}\omega_{130}}}  - (r^2-1) \sqrt{\frac{\omega_{110}\omega_{130}}{\omega_{011}^2}} \right),\label{22-2}\\
&\expval{(\vb B\vb E)^2} = \frac{4\kappa}{2\sqrt{2}V}\frac{\pi^2}{L_z^2} r^2\left(3\sqrt{\frac{\omega_{011}^2}{\omega_{110}\omega_{130}}} +3 \sqrt{\frac{\omega_{110}}{\omega_{130}}} -  \sqrt{\frac{\omega_{130}}{\omega_{110}}} - \sqrt{\frac{\omega_{110}\omega_{130}}{\omega_{011}^2}}  \right).
\label{22-3}
\end{align}
%Taking into account the expressions (\ref{eq:free-3d}) for given mode frequencies
%$$
%\omega_{011} = \frac{\pi}{L_z}\sqrt{r^2 + 1}, \qquad \omega_{110} = \frac{\pi}{L_z}\sqrt{2}\,r, \qquad \omega_{130} = \frac{\pi}{L_z}\sqrt{10}\,r,
%$$
Collecting all contributions (\ref{22-1})--(\ref{22-3}) together and taking into account the expressions (\ref{energycons}) and (\ref{omegas}), one obtains,
\begin{equation}
    {\sf M}_{2\to 2} = \frac{-\kappa}{L_z^5} \frac{ (2\pi)^2 r^3}{\sqrt[4]{5} \sqrt{1+r^2}} \qty[5 + 2\sqrt{5} - \beta \qty(\sqrt{1+r^2} + \sqrt{2}r)^2].
    \label{M22}
\end{equation}
The probability for the $2 \to 2$ process is a squared modulo of the matrix element, including the restored energy delta-function part. Following the dissipation arguments from Sec.~3, the delta-function part %to be replaced with
should be interpreted as the mode minimal quality factor, which decreases with the mode frequency for SRF cavity \cite{1062561}. Precisely, we replace $2\pi\delta\left(\Sigma\omega_i - \Sigma\omega_f\right)$ with $Q/\omega_{130}$. Finally,
%Restoring the S-matrix element $\int d t\,e^{it(2\omega_{011}-\omega_{110}-\omega_{130})}= 2\pi\delta(2\omega_{011} - \omega_{110} - \omega_{130})$, the probability for the elementary $2 \to 2$ process reads, 
\begin{equation}
    \label{eq:mel}
    P_{2 \to 2} = \abs\big{\mel{f}{\sf S}{i}}^2 =% \qty[2\pi\delta(0)]^2 
    G_1^2 \frac{\kappa^2 Q^2}{L_z^{8}}, \quad  G_1^2 = \frac{2^3 \pi^2 r^4}{5^{3/2} (1+r^2)} \qty[5 + 2\sqrt{5} - \beta \qty(\sqrt{1+r^2} + \sqrt{2}r)^2]^2.
\end{equation}
%Here $2\pi\delta(0)$ should be replaced to the minimal quality factor for modes, which decreases with the mode frequency for SRF cavity \cite{1062561}, so we replace  $2\pi\delta(0) \to Q/\omega_{130}$. 

This indicates that a single cavity pump mode can generate two signal modes if both of them are resonant. This process may be potentially probed experimentally. However, the effect of Bose enhancement shows that if one of the signal modes is already excited at the $n$-th level ($n$ quanta), the corresponding amplitude will be enhanced in $\sqrt{n}$ times. So, we come to the case of two pump modes implying the ``minus'' signal mode generation.

%Bose enhancement 

%However, for ``minus'' signal mode generation one should consider coherent states.

\subsection{``Minus'' signal mode generation by the coherent states}

In this subsection we apply the $2 \to 2$ matrix element to the generation of one-particle state of the signal mode  $ \ket{1^\text{TM}_{130}}$  in the field of two coherent states of pump modes. Here we will focus on a small mean number of signal mode quanta $\expval{N_s} \lesssim 1$ in the steady regime, so we neglect the processes with two and more quanta of signal mode in the final state for simplicity\footnote{In general, one should take into account many-particle states of the signal mode as well.}. 
%Instead of pump modes of large amplitudes described by coherent states, we expect  generation of a very small number of signal photons. Hence, we consider one-particle final state $ \ket{1^\text{TM}_{130}}$ for the signal mode.
Precisely, the initial and final states read, 
%A solution to this %(illusory? discrepancy % (controversy? disharmony?) is to introduce \emph{coherent} pump modes, which drive %(enable? launch? make possible?) the scattering process.
\begin{align*}
    \label{eq:states-coh}
    \ket{i} &= \ket{\xi^\text{TE}_{011}} \otimes \ket{\eta^\text{TM}_{110}} = e^{-\frac{|\xi|^2+|\eta|^2}{2}} \sum\limits_{i,j = 0}^\infty \frac{\xi^i \eta^j}{i! j!} \qty(a^{\rm TE+}_{011})^i \qty(a^{\rm TM+}_{110})^j \ket{0},
    \\
    \ket{f} &= \ket{\xi^\text{TE}_{011}} \otimes \ket{\eta^\text{TM}_{110}} \otimes \ket{1^\text{TM}_{130}} = \ket{i} \otimes \ket{1^\text{TM}_{130}} = a^{\text{TM}+}_{130} \ket{i}.
\end{align*}
%The choice of coherent states rather than pure ones is also a more ``physical'' choice, as long as coherent state is a closer analog of classical wave of a large amplitude. 
The parameters $\xi, \eta$ are associated with the mean number of quanta in the pump modes,
\begin{equation}
    \expval{ N_{\text{TE}_{011}} } = |\xi|^2, \qquad  \expval{ N_{\text{TM}_{110}} } =  |\eta|^2 .
    \label{Nxieta}
\end{equation}
%Note that we expect generation of single signal photons TM130 --- therefore a pure-state ket-vector $\ket{1^\text{TM}_{130}}$ is chosen.
The matrix element related to the coherent generation connected to the $2 \to 2$ matrix element as,
\begin{equation}
\label{eq:mel-coh}
 {\sf M}_\text{coh} =    e^{-|\xi|^2-|\eta|^2} \int_V d^3x  \sum\limits_{i,j,k,l=0}^\infty \frac{\xi^i \eta^j}{i!j!} 
\times\frac{\qty(\xi^\ast)^k \qty(\eta^\ast)^l}{k!l!} \expval{a^{\text{TM}-}_{130} \qty(a^{\text{TE}-}_{011})^k \qty(a^{\text{TM}-}_{110})^l \, \nord{{\cal L}_{\rm int}} \, \qty(a^{\text{TE}+}_{011})^i \qty(a^{\text{TM}+}_{110})^j}{0},
\end{equation}
where ${\cal L}_{\rm int}$ determined in (\ref{eq:EH-term}). Performing the Wick contractions similarly to that done in Sec.3.2., we obtain the matrix element (\ref{M22}) with additional combinatorial factor $li(i-1)$ and the factor $ (l-1)! (i-2)! \delta_{l-1,j} \delta_{k,i-2}$ from remaining after the Wick contraction field operators in the initial and final states, yielding
\begin{equation}
 {\sf M}_\text{coh} = e^{-|\xi|^2-|\eta|^2} \sum\limits_{i,j,k,l=0}^\infty \frac{\xi^i \eta^j}{j!} 
\times\frac{\qty(\xi^\ast)^k \qty(\eta^\ast)^l}{k!} \delta_{l-1,j} \delta_{k,i-2} \times \sqrt{2}\,M_{2\to 2} = \xi^2 \eta^*  \times  \sqrt{2}\, {\sf M}_{2\to 2}.
\end{equation}
Therefore, the probability for the coherent process reads (recall  \eqref{Nxieta}),
\begin{equation}
\label{Pcoh}
   P_\text{coh} = 2\expval{N_{\text{TE}_{011}}} ^2 \cdot \expval{N_{\text{TM}_{110}}} \cdot P_{2\to 2}.
\end{equation}
We want to emphasize the natural appearance of a Bose enhancement factor $\expval{N_{\text{TM}_{110}}}$ that relates to the process initiated by a single pump mode TE$_{011}$.

The transition probability (\ref{Pcoh}) can be interpreted as the mean steady number of signal quanta in the final state. 
%The mean steady number of signal mode quanta can be connected with the transition probability with the Keldysh formalism. 
Precisely, averaging the operator of the number of signal mode quanta $\hat{N}_s \equiv  a^\dag_{s}a_{s}$ (for our choice of modes $a_{s}=a_{\rm TM_{130}}$) over the initial state $\ket{i}$ includes the time evolution from $t_i$ to $t_f$ and back,
\begin{equation}   
\expval{N_s}=\mel{i}{\hat{N}_{s}}{i} = \int df' df'' \mel{i}{S^\dag}{f'} \mel{ f'}{\hat{N}_{s}}{f''} \mel{f''}{S}{i}.
\end{equation}
Here $\ket{f'}$ and $\ket{f'}$ represent the full number of states in theory at the moment $t_f$. We suppose that the pump mode coherent states remain unchanged; the only difference are the excited states of the signal mode. Finally, only $\ket{f'}=\ket{f''}=\ket{f}$ give nonzero outcome, so we have $\expval{N_s} = P_\text{coh}$. 

%\begin{equation}
%    \label{eq:n-quant-pre}
%     \langle N_s \rangle = G_1^2 \frac{\kappa^2Q^2|\xi|^4|\eta|^2}{L_z^8}, \qquad G_1^2 = \frac{\tilde G_1^2}{10\pi^2r^2}.
%\end{equation}
%Finally, the mean numbers of pump photons $\abs{\xi}^2$ and $\abs{\eta}^2$ should be expressed via classical field amplitudes,
%\begin{align*}
%    \label{eq:convert-amps}
%    |\xi|^2 &= N^\text{TE}_{011} = \int_V\frac{\vb E^2_{011} + \vb B^2_{011}}{2\omega_{011}}\dd^3 x = \frac{B_0^2V}{2\omega_{011}} = \frac{B_0^2 L_z^4}{2\pi r^2\sqrt{1+r^2}},
%    \\
%    |\eta|^2 &= N^\text{TM}_{110} = \int_V\frac{\vb E^2_{110} + \vb B^2_{110}}{2\omega_{110}}\dd^3 x = \frac{B_0^2V}{2\omega_{110}} = \frac{B_0^2 L_z^4}{2\sqrt{2} \pi r^3}.
%\end{align*}
%Substitution of these into eq. \eqref{eq:n-quant-pre} results in a complete QFT prediction of the mean steady number of photons, constituting the signal field:

Remind the connection between the mean photon number and the classical amplitude % $\expval{N_{p}} = \int_V \dd^3 x (\vb E^2_{p} + \vb B^2_{p})/(2\omega_{p}) = F_0^2V/(2\omega_{p})$
\begin{equation}
\label{N-ampl}
\expval{N_{p}} = \int_V\frac{\vb E^2_{p} + \vb B^2_{p}}{2\omega_{p}}\dd^3 x = \frac{F_0^2V}{2\omega_{p}},
\end{equation}
where $F_0$ is the amplitude for electric and magnetic field  for each pump mode.   Finally, one obtains
\begin{equation}
    \label{eq:n-quant}
   \expval{N_s} =\tilde{G}_1^2 \times \kappa^2 Q^2 F_0^6 L_z^4, \qquad  \tilde G_1^2 = \frac{4}{(10)^{3/2}\pi r^3(1+r^2)^2} \qty[5 + 2\sqrt{5} - \beta \qty(\sqrt{1+r^2} + \sqrt{2}r)^2]^2.
\end{equation}
%Comparing this quantity with its classical counterpart, calculated with the classical amplitude given in \cite{OurArticle}, we merrily discover their exact equality. % we enjoy their exact coincidence.
 Numerically, one obtains $\tilde{G}_1^2 = 3.3$ for $r=\sqrt{\sqrt{5}-2}$ and $\beta = 7/4$.
The result (\ref{eq:n-quant}) coincides in terms of the mean number of signal quanta (\ref{N-ampl}) with one obtained by solving classical equation in the effective field theory approach \cite{OurArticle}.

\subsection{Towards experimental probes}

Alghough the article is mostly devoted to the details of the matrix element calculation responcible to the signal mode generation, it is worth to review the perspectives of experimental detection of the nonlinear effect. 
 We follow \cite{Eriksson:2004cz,Bogorad:2019pbu} and experimental proposals mentioned in \cite{Giaccone:2022pke, Kahn:2022uko}. 

The main issue of experimental detection is to discriminate the signal events from the thermal noise. For this reason the Dicke radiometer equation is used in the literature  \cite{Eriksson:2004cz,Bogorad:2019pbu},
\begin{equation}
SNR = \frac{P_s}{T} \sqrt{\frac{t}{B}}= \frac{\langle N_s \rangle}{T}\frac{\omega_s}{L_z Q}\sqrt{\frac{t}{B}},
\end{equation}
Here $P_s$ is the signal power,  $T$ --- the cavity temperature, $t$ --- is the total measurement time, and $B$ is the bandwidth which can be taken as $B = 1/t$. Fixing SNR as a characteristic of a given detector, one inverses the equation (\ref{eq:n-quant}), and obtain the necessary measuring time necessary to detect  nonlinearity with the coupling constant $\kappa$,
%\begin{equation}
%\label{sensitivity}
%\kappa =  \left( \frac{SNR\,\cdot T}{\tilde{G}_1^2Q F_0^6 L_z^3 \omega_s} {\frac{1}{t}} \right)^{1/2},
%\end{equation}
\begin{equation}
\label{sensitivity}
t =   \frac{SNR\,\cdot T}{\tilde{G}_1^2Q F_0^6 L_z^3 \omega_s \kappa_{EH}^2}\left(\frac{\kappa_{EH}}{\kappa} \right)^2 ,
\end{equation}
where $\kappa_{EH} = \frac{\alpha_{em}^2}{90m_e^4}$ related to  the Euler-Heisenberg theory. Let us estimate the meausuring time (\ref{sensitivity}) numerically. Assuming $L_z = 20$ cm (hence $\omega_s = 2.4\cdot 10^{-6}$ eV), $F_0 = 0.1$ T, $Q = 10^{10}$ and $T = 1$ K, $\mathrm{SNR} \simeq 5$, one can estimate that the Euler-Heisenberg interaction can be detected in $22$ seconds. 

Additional issues include the filtering the signal mode from the pump modes. For this reason the authors of Ref.~\cite{Bogorad:2019pbu}  proposed the special filtering cavity bottle-shape geometry in which only the signal penetrates into a bottleneck. The large values for the pump mode field strength can be supported by the cavity walls since the magnetic field strength of pump modes is supposed to be less than $0.2$ T which is the critical magnetic field for cavity walls made from pure Niobium.

%Assuming $L_z = 20$ cm, $F_0 = 0.2$ T, $Q = 10^{11}$ and $T = 50$ mK one estimates the $\mathrm{SNR} \simeq 4$ (while $\expval{N_s} \simeq 20$) if coherence time is $t_\text{coh} = Q/\omega_s \simeq 14$ s and $B = 1/t$. % Despite tractable SNR, 
%the noise contribution from superconductor nonlinearities in cavity walls is still to be examined and is beyond the scope of this paper.

%An open question is to study the possibility for generating nonlinear
  % discriminating the signal events from the thermal and other noises and taking into account the nonlinearities in the cavity walls.

\section{Discussion}
\label{sec:conclude}

We applied the technique of quantum perturbative calculations for electromagnetic field with nonlinear 4-photon interaction in a closed cavity. We calculated the amplitudes and probabilities of $3 \to 1$ merging and $2 \to 2$ scattering elementary processes.

We clarify the following points which remain unclear in the classical effective theory calculation \cite{OurArticle}. First, %(the explanation of)
the matrix elements for the third harmonics generation and ``plus'' combination modes vanish since the matrix elements in plane wave decomposition contain only scalar products of momenta which vanish in presence of Lorentz invariance.

Secondly, we provide a clear quantum interpretation for the signal mode ``minus'' generation by an interaction of two pump modes. Thus, it turns out that the crucial elementary process is $2 \to 2$ process in terms of quanta. The overall amplitude is nonzero since the coherent states which describe classical waves contain a fluctuating number of quanta so the whole number does not have to conserve. We have shown that the same elementary process is responsible for two signal modes generation by a single pump mode; the process 
driven by two pump modes is the same including Bose enhancement.

The developed formalism of perturbative calculations of transition probabilities between cavity states may be generalized to other choices of initial and final states, which may include, for example, squeezed states etc. One more way of generalization is to consider average values of various operators, correlators (see Schwinger-Keldysh formalism for an interacting scalar field in one-dimensional cavity \cite{Trunin:2022dvg}).
An application of the formalism to cavity processes including $4$-photon interaction in a nonlinear media may be also interesting.

%Making an interpretation  

%More accurate quantum calculation may include the calculation of several operators in Keldysh-Schwinger formalism, dissipation may be included as ... formalism.   

%Alternative --- operator equation in Heisenberg representation
%This is out of scope of the current paper 

\paragraph{Acknowledgments} The Authors thank Maxim Fitkevich, Dmitry Kirpichnikov, Dmitry Levkov, \\ \fbox{Valery Rubakov}, Alexey Rubtsov and Dmitry Salnikov for helpful discussions. The work is supported by RSF grant 21-72-10151.

\appendix
\section {Elements of calculation of the matrix element for the \texorpdfstring{$3 \to 1$}{3 --> 1} merging process}
\label{app:details-3-1}

In this Appendix we provide more details for the merging matrix element calculation \eqref{Tfi}, whose expansion into the components of electromagnetic field strength operators has the following form, 
\begin{align}
    \nonumber
   \mel{f}{\sf S }{i}  &= 4i\kappa S \int\limits_{-\infty}^{+\infty}\dd t \int\limits_0^{L_x}\dd x\expval{a^-_{l,3n} \nord{\vb E^4 - 2\vb B^2 \vb E^2 + \vb B^4 + 4\beta(\vb B\vb E)^2} \qty(a^+_{i,n})^2 a^+_{j,n}}{0} =
    \\ \label{eq:1d-mel}
    &= \expval{\vb E^4} - 2\expval{\vb B^2\vb E^2} + \expval{\vb B^4} + 4\beta\expval{(\vb B\vb E)^2} =
    \\ \nonumber
    &= \qty\Big{\expval{E_y^4} + 2\expval{E_y^2E_z^2} + \expval{E_z^4}} ~-~ 2\qty\Big{\expval{B_y^2 E_y^2} + \expval{B_y^2 E_z^2} + \expval{B_z^2 E_y^2} + \expval{B_z^2 E_z^2}}
    \\ \nonumber
    &+~ \qty\Big{\expval{B_y^4} + 2\expval{B_y^2B_z^2} + \expval{B_z^4}} ~+~ 4\beta\qty\Big{\expval{B_y^2 E_y^2} + 2\expval{B_y E_y B_z E_z} + \expval{B_z^2 E_z^2}}.
\end{align}
Here the brackets for the field strength operators are defined in the first line of \eqref{eq:1d-mel}. In total, $13$ terms appear in eq.~\eqref{eq:1d-mel}. %arise %appear.
Let us present an evaluation for one of them:
\begin{align*}
\expval{E_y^4} &= 4i\kappa S \int\limits_{-\infty}^{+\infty}\dd t \int\limits_0^{L_x}\dd x \wick[below]{\expval{\c1 a^-_{l,3n} \,\nord{\c1 E_y\c3 E_y\c2 E_y\c1 E_y}\, \c1 a^+_{i,n} \c2 a^+_{i,n} \c3 a^+_{j,n}}{0}} =
 %   \expval{E_y^4} &= 4i\kappa S \int\limits_{-\infty}^{+\infty}\dd t \int\limits_0^{L_x}\dd x \wick[below]{\expval{\c1 a^-_{l,3n} \,\nord{\c1 E_y\c3 E_y\c2 E_y\c1 E_y}\, \c1 a^+_{i,n} \c2 a^+_{i,n} \c3 a^+_{j,n}}{0}} =
    \\
    &= 4i\kappa S \int\limits_{-\infty}^{+\infty}\dd t \int\limits_0^{L_x}\dd x \,4!\, \qty(+i\delta_{ly} \sqrt{\frac{\omega_{3n}}{V}} \sin(k_{3n}x) \, e^{+i\omega_{3n}t}) ~\times
    \\
    &\qquad\qquad \times~ \qty(-i\delta_{iy} \sqrt{\frac{\omega_n}{V}} \sin(k_nx) \, e^{-i\omega_nt})^2 \qty(-i\delta_{jy} \sqrt{\frac{\omega_n}{V}} \sin(k_nx) \, e^{-i\omega_nt}) =
    \\
    &= 4i\kappa S \,4!\, \delta_{iy}\delta_{jy}\delta_{ly} \, i(-i)^3 \, \sqrt{\frac{\omega_{3n}\omega_n^3}{V^4}} \int\limits_{-\infty}^{+\infty} e^{i(\omega_{3n}-3\omega_n)t} \dd t \int\limits_0^{L_x}\sin(k_{3n}x)\sin^3(k_nx)\dd x =
    \\
    &= 2\pi\delta(0) \, \frac{12\sqrt{3}i\pi^2 n^2\kappa}{L_x^3 S} \, \delta_{iy}\delta_{jy}\delta_{ly}.
\end{align*}
%Данный расчёт показывает, что традиционно возникает $\delta$-функция, выражающая закон сохранения энергии в процессе, а четыре собственные функции после интегрирования по $x$ собираются в геометрический множитель. Отдельно стоит обратить внимание на комбинаторный коэффициент $\boxed{4!}$, возникающий в силу неразличимости полей в произведении $\nord{E_y^4}$. При усреднении, например, $\expval{B_y^2 E_y^2}$ комбинаторный множитель составит $2! \cdot 2! = 4$, а член $\expval{B_y E_y B_z E_z}$ даст единичный комбинаторный фактор.
Here $4!$ is a combinatorial factor arising %appeared
due to the identity of four $E_y$ operators. At the last step we use $\omega_n=k_n=\frac{\pi n}{L_x}$ and the similar relation for the $3n$ subscript.  

\section{Details of the matrix element \texorpdfstring{$2 \to 2$}{2 --> 2} calculation}
\label{app:details-2-2}
Let us present the calculation for the terms of the matrix element \eqref{eq:mel-2-2} in more detail. The first term $\expval{ (\vb E\vb E)^2}$ reads,
\begin{align}
\expval{ (\vb E\vb E)^2}: \qquad \qquad &\int\dd^3{\rm x}  \mel{0}{\tfrac{1}{\sqrt{2}}a^{\text{TE}}_{110}a^{\text{TE}}_{110}\,\nord{\vb{ (EE)(EE)}}\,a^{\text{TM}+}_{110}a^{\text{TM}+}_{130}}{0}  =
   % \\
     (+i)^2(-i)^2 \frac{\sqrt{\omega_{130}\omega_{011}^2\omega_{110}}}{\sqrt{2}\sqrt{2^4V^4}}\times \notag
    \\
    \times &\int\dd^3{\rm x} \,8\Bigl[\underbrace{(\vb*{\mathcal  A}^\text{TM}_{130}\vb*{\mathcal A}^\text{TM}_{110}) (\vb*{\mathcal A}^\text{TE}_{011}\vb*{\mathcal A}^\text{TE}_{011})}_{\int = -V/2} ~+~ 2 \underbrace{(\vb*{\mathcal A}^\text{TM}_{130}\vb*{\mathcal A}^\text{TE}_{011}) (\vb*{\mathcal A}^\text{TM}_{110}\vb*{\mathcal A}^\text{TE}_{011})}_{\int = 0}\Bigr] = \frac{-\sqrt{\omega_{130}\omega_{011}^2\omega_{110}}}{\sqrt{2}V}. 
    \label{eq:e4-2-2}
\end{align}
Here 8 and 2 are combinatorial coefficients, for shortness we indicate the result of integration with underbraces. Thus, the second term in \eqref{eq:e4-2-2} vanishes.  

Analogically, the following term $\expval{ (\vb B\vb B)^2}$ reads,
\begin{align}
\expval{ (\vb B\vb B)^2}: \qquad \int\dd^3{\rm x}  \mel{0}{\tfrac{1}{\sqrt{2}}a^{\text{TE}}_{110}a^{\text{TE}}_{110}\,\nord{\vb{ (BB)(BB)}}\,a^{\text{TM}+}_{110}a^{\text{TM}+}_{130}}{0}  =
   % \\
     (+i)^2(-i)^2 \frac{1}{\sqrt{2}\sqrt{2^4V^4}\sqrt{\omega_{130}\omega_{011}^2\omega_{110}}}\times \notag
    \\
    \times \int\dd^3{\rm x} \,8\Bigl[\underbrace{(\rm rot \vb{\mathcal  A}^\text{TM}_{130}\rm rot\vb{\mathcal A}^\text{TM}_{110}) (\rm rot\vb{\mathcal A}^\text{TE}_{011}\rm rot\vb{\mathcal A}^\text{TE}_{011})}_{\int = V/2\cdot \pi^4(L_{x}^{-2} + 3 L_{y}^{-2}) (L_{y}^{-2} - L_{z}^{-2})} ~+~ 2 \underbrace{(\rm rot\vb*{\mathcal A}^\text{TM}_{130}\rm rot\vb*{\mathcal A}^\text{TE}_{011}) (\rm rot \vb*{\mathcal A}^\text{TM}_{110}\rm rot\vb*{\mathcal A}^\text{TE}_{011})}_{\int = -V/2 \cdot \pi^4 \,L_{x}^{-2} L_{z}^{-2}}\Bigr] =\\
    = \frac{\pi^4}{\sqrt{2}V}\frac{(L_{x}^{-2} + 3 L_{y}^{-2}) (L_{y}^{-2} - L_{z}^{-2}) - 2 L_{x}^{-2} L_{z}^{-2}}{\sqrt{\omega_{130}\omega_{011}^2\omega_{110}}} = \frac{1}{\sqrt{2}V} \frac{\pi^4}{L_z^4}2r^2\frac{ 2r^2 - 3}{ \sqrt{\omega_{011}^2\omega_{110}\omega_{130}}}. \notag
\end{align}
The combinatorial coefficients $8$ and $2$ coincide with the previous case. However, the result of the integration differs. The first of the mixed terms $\expval{ (\vb B\vb B) (\vb E\vb E)}$ reads,
\begin{align}
 	\expval{ (\vb B\vb B) (\vb E\vb E)}: \qquad \qquad \qquad  \int\dd^3{\rm x}  \mel{0}{\tfrac{1}{\sqrt{2}}a^{\text{TE}}_{110}a^{\text{TE}}_{110}\,\nord{\vb{ (BB)(EE)}}\,a^{\text{TM}+}_{110}a^{\text{TM}+}_{130}}{0}  =
	\frac{1}{\sqrt{2^4V^4}} \, \int\dd^3{\rm x}\, 4\,\times \notag
	\\
	\times \Biggl[(-i)^2\sqrt{\frac{\omega_{011}^2}{\omega_{130}\omega_{110}}}\underbrace{(\rm rot\vb{\mathcal A}^\text{TM}_{130}\rm rot\vb{\mathcal A}^\text{TM}_{110})(\vb{\mathcal A}^\text{TE}_{011}\vb{\mathcal A}^\text{TE}_{011})}_{\int = -V/2 \cdot \pi^2(L_{x}^{-2} + 3L_{y}^{-2})}
	~
	+~ i^2\sqrt{\frac{\omega_{130}\omega_{110}}{\omega_{011}^2}}\underbrace{(\rm rot\vb{\mathcal A}^\text{TE}_{011}\rm rot\vb{\mathcal A}^\text{TE}_{011}) (\vb{\mathcal A}^\text{TM}_{130}\vb{\mathcal A}^\text{TM}_{110})}_{\int = V/2 \cdot \pi^2(L_{y}^{-2} - L_{z}^{-2})}
	~+
	\\
	+~2i(-i)\sqrt{\frac{\omega_{110}}{\omega_{130}}}\underbrace{(\rm rot\vb{\mathcal A}^\text{TM}_{130}\rm rot\vb{\mathcal A}^\text{TE}_{011}) (\vb{\mathcal A}^\text{TM}_{110}\vb{\mathcal A}^\text{TE}_{011})}_{\int = 0} ~+
	~ 2i(-i)\sqrt{\frac{\omega_{130}}{\omega_{110}}}\underbrace{(\rm rot\vb{\mathcal A}^\text{TM}_{110}\rm rot\vb{\mathcal A}^\text{TE}_{011}) (\vb{\mathcal A}^\text{TM}_{130}\vb{\mathcal A}^\text{TE}_{011})}_{\int = 0}\Biggr] =  \notag
	\\
	= \frac{1}{2\sqrt{2}V}\frac{\pi^2}{L_z^2}\left(4r^2 \sqrt{\frac{\omega_{011}^2}{\omega_{110}\omega_{130}}}  - (r^2-1) \sqrt{\frac{\omega_{110}\omega_{130}}{\omega_{011}^2}} \right). \notag
\end{align}
The last term,
\begin{align}
 	\expval{ (\vb B\vb E)^2 }: \qquad \qquad \qquad  \int\dd^3{\rm x}  \mel{0}{\tfrac{1}{\sqrt{2}}a^{\text{TE}}_{110}a^{\text{TE}}_{110}\,\nord{\vb{ (BE)(BE)}}\,a^{\text{TM}+}_{110}a^{\text{TM}+}_{130}}{0}  =
	\frac{1}{\sqrt{2^4V^4}} \, \int\dd^3{\rm x}\, 4\,\times \notag
	\\
	\times \Biggl[(-i)^2\sqrt{\frac{\omega_{011}^2}{\omega_{130}\omega_{110}}}\underbrace{(\rm rot\vb{\mathcal A}^\text{TM}_{130}\vb{\mathcal A}^\text{TE}_{011})(\rm rot\vb{\mathcal A}^\text{TM}_{110}\vb{\mathcal A}^\text{TE}_{011})}_{\int = -3V/2 \cdot \pi^2 L_{y}^{-2}}
	~
	+~ i^2\sqrt{\frac{\omega_{130}\omega_{110}}{\omega_{011}^2}}\underbrace{(\rm rot\vb{\mathcal A}^\text{TE}_{011}\vb{\mathcal A}^\text{TM}_{130}) (\rm rot\vb{\mathcal A}^\text{TE}_{011}\vb{\mathcal A}^\text{TM}_{110})}_{\int = V/2 \cdot \pi^2 L_y^{-2}}
	~+
 \\
 +~i(-i)\sqrt{\frac{\omega_{110}}{\omega_{130}}}\underbrace{(\rm rot\vb{\mathcal A}^\text{TM}_{130}\vb{\mathcal A}^\text{TE}_{011}) (\rm rot\vb{\mathcal A}^\text{TM}_{110}\vb{\mathcal A}^\text{TE}_{011})}_{\int = 0} ~+
	~ i(-i)\sqrt{\frac{\omega_{130}}{\omega_{110}}}\underbrace{(\rm rot\vb{\mathcal A}^\text{TM}_{110}\vb{\mathcal A}^\text{TM}_{130}) (\rm rot\vb{\mathcal A}^\text{TE}_{011}\vb{\mathcal A}^\text{TE}_{011})}_{\int = 0}+   \notag
	\\
	+~i(-i)\sqrt{\frac{\omega_{110}}{\omega_{130}}}\underbrace{(\rm rot\vb{\mathcal A}^\text{TM}_{130}\vb{\mathcal A}^\text{TE}_{011}) (\rm rot \vb{\mathcal A}^\text{TE}_{011}\vb{\mathcal A}^\text{TM}_{110})}_{\int = 3V/2\cdot \pi^2 L_y^{-2}} ~+
	~ i(-i)\sqrt{\frac{\omega_{130}}{\omega_{110}}}\underbrace{(\rm rot\vb{\mathcal A}^\text{TM}_{110}\vb{\mathcal A}^\text{TE}_{011}) (\rm rot\vb{\mathcal A}^\text{TE}_{011}\vb{\mathcal A}^\text{TM}_{130})}_{\int = 0}\Biggr] =  \notag
	\\
	=\frac{1}{2\sqrt{2}V}\frac{\pi^2}{L_z^2} r^2\left(3\sqrt{\frac{\omega_{011}^2}{\omega_{110}\omega_{130}}} +3 \sqrt{\frac{\omega_{110}}{\omega_{130}}} -  \sqrt{\frac{\omega_{130}}{\omega_{110}}} - \sqrt{\frac{\omega_{110}\omega_{130}}{\omega_{011}^2}}  \right). \notag
\end{align}
Finally, restoring the coefficient $4\kappa$, the result \eqref{22-1}--\eqref{22-3} is obtained.
\bibliographystyle{unsrt}
\bibliography{biblio}

\end{document}